\documentclass{sig-alternate-05-2015}

\usepackage{graphicx}
\usepackage{epsfig}
\usepackage{enumitem}
\usepackage{amsmath}

\begin{document}

\title{SAT-based Analysis of\\ Large Real-world Feature Models is Easy}

\numberofauthors{4}
\author{
\alignauthor
Jia Hui Liang \\
       \affaddr{University of Waterloo, Canada}
\alignauthor
Vijay Ganesh \\
       \affaddr{University of Waterloo, Canada}
\and
\alignauthor
Krzysztof Czarnecki \\
       \affaddr{University of Waterloo, Canada}
\alignauthor
Venkatesh Raman \\
       \affaddr{Institute of Mathematical Sciences, India}
}
       

\setcopyright{acmcopyright}
\conferenceinfo{SPLC 2015,}{July 20 - 24, 2015, Nashville, TN, USA}
\isbn{978-1-4503-3613-0/15/07}\acmPrice{\$15.00}
\doi{http://dx.doi.org/10.1145/2791060.2791070}

\maketitle
\begin{abstract}
  \sloppypar{Modern conflict-driven clause-learning (CDCL) Boolean SAT
    solvers provide efficient automatic analysis of
    real-world feature models (FM) of systems ranging from cars to
    operating systems. It is well-known that solver-based analysis of
    real-world FMs scale very well even though SAT instances obtained
    from such FMs are large, and the corresponding analysis problems
    are known to be NP-complete. To better understand why SAT solvers
    are so effective, we systematically studied many syntactic and
    semantic characteristics of a representative set of large
    real-world FMs. We discovered that a key reason why large
    real-world FMs are easy-to-analyze is that the vast majority of
    the variables in these models are {\it unrestricted}, i.e., the
    models are satisfiable for both true and false assignments to such
    variables under the current partial assignment. Given this discovery and our understanding of
    CDCL SAT solvers, we show that solvers can easily find satisfying
    assignments for such models without too many backtracks relative
    to the model size, explaining why solvers scale so well. Further
    analysis showed that the presence of unrestricted variables in
    these real-world models can be attributed to their high-degree of
    {\it variability}. Additionally, we experimented with a series of
    well-known non-backtracking simplifications that are particularly
    effective in solving FMs. The remaining
    variables/clauses after simplifications, called the {\it
      core}, are so few that they are easily solved even with backtracking,
      further strengthening our
    conclusions. In contrast to our findings, previous research
    characterizes the difficulty of analyzing randomly-generated
    FMs in terms of treewidth. Our experiments suggest that the
    difficulty of analyzing real-world FMs cannot be explained in
    terms of treewidth.}
\end{abstract}




\section{Introduction}
Feature models (FM) are widely used to represent the variablility and
commonality in product lines and reusable software, first introduced
in 1990~\cite{Kang1990}. Feature models define all the valid feature
configurations of a product line, whether it be a car or software. The process of feature modeling helps to ensure that
relevant features are reusable, and unused features are removed to
lower the complexity of the system under
design~\cite{GenerativeProgramming}. Promoting the reuse of features
shortens the time of delivering new products and lowers the cost of
production. Such models can be automatically analyzed using
conflict-driven clause-learning (CDCL) Boolean SAT solvers to detect inconsistencies or errors in the design
process of a product, thus lowering the cost of production.

Modern CDCL Boolean SAT solvers are known to
efficiently solve many large real-world SAT instances obtained from a
variety of domains such as software testing, program analysis, and
hardware verification. More recently, inspired by the success of SAT
solvers in other domains, many researchers proposed the use of solvers
to analyze feature
models~\cite{BatorySPLC,BenavidesSurvey,BenavidesVAMOS}. Subsequently,
solvers have been widely applied to perform all manner of analysis on
feature models, and have proven to be surprisingly effective even
though the kind of feature model analysis discussed here is an NP-complete problem.
Furthermore, real-world FMs tend to be
very large, often running into hundreds of thousands of clauses and
tens of thousands of variables. This state of affairs has perplexed
practitioners and theoreticians alike.

\noindent{\bf Problem Statement:} Hence, the problem we address in
this paper is ``Why is CDCL SAT-based analysis so effective in analyzing large real-world FMs?''

\noindent{\bf Contributions:} Here we describe the
contributions made in this paper.

\begin{enumerate}[leftmargin=*]

\item We found that the overwhelming
number of variables occuring in real-world FMs (or its equivalent
Boolean formula) that we studied are {\it unrestricted}. We say that a
variable $v$ in a Boolean formula $\phi$ given partial assignment $S$
is unrestricted if there exist two satisfying extensions of $S$ (i.e.,
extensions of the partial assignment of $S$ to all variables in $\phi$
such that $\phi$ is satisfiable), one with $v=true$ and the other with
$v=false$. Intuitively, an unrestricted variable does not cause modern
CDCL SAT solvers to backtrack because under the given partial
assignment, the solver cannot assign the unrestricted variable to a wrong value.
This is important
because if the number of backtracks a solver performs remains small
relative to the size of inputs, then the solver is likely
to scale very well for such class of inputs. Indeed, this is exactly
what we observed in our experiments, CDCL SAT solvers perform
very few, near-constant, number of backtracks even as the size of FMs
increase.
Prior to our findings, there was no known reason to believe that a
majority of variables in real-world FMs are unrestricted.
Note that in CDCL SAT solvers,
the number of backtracks can be worst-case exponential in the number
of Boolean variables. Hence, our finding of the link between large number
of unrestricted variables
in real-world FMs and near-constant number of backtracks by CDCL
solvers in solving such instances goes to the heart of the question
posed in the paper. Modern CDCL SAT solvers contain many features
that improve their performance. We found that switching off these features,
excluding Boolean constraint propagation (BCP) and
backjumping, had no negative impact on performance while solving FMs. This observation
is consistent with the fact that the vast majority of variables in
real-world FMs are unrestricted.

\item We also investigated the possible source of unrestricted
  variables in real-world FMs. We observed that the large percentage
of unrestricted variables in real-world FMs is attributable to the
{\it high variability} in such models. We say that an FM has high
variability if a large number of features occuring in such a model
are optional, i.e., one can obtain a valid product configuration
irrespective of whether such features are selected. Indeed, this
observation is consistent with the fact that FMs capture all
configurations of a product line, whereas deriving a valid product from such
models only require relatively few features to be present.

\item We implemented numerous
  well-known non-backtracking simplifications
  from SAT literature. These techniques are invoked as a pre-processing step, prior
  to calling the solver. Often these simplifications completely solve
  such instances. There are few instances where these simplifications
  did not completely solve the input FMs, and instead returned a very
  small simplified formula, we call a {\it core}. These cores tend to
  consist largely of {\it Horn} clauses, that are subsequently easily
  solved by solvers with near-constant number of backtracks in the
  worst-case.
  The point of these experiments was not to suggest new techniques,
  but to further test our hypothesis that SAT-based analysis of FM is easy
  by demonstrating the effectiveness of polynomial-time simplifications
  on the FM inputs. The main simplifications are based on resolution which
  is the basis of CDCL solvers. Efficient implementations
  of such simplification techniques are part of many modern
  CDCL SAT solvers such as Lingeling or Glucose.
  

\item Following previous work by Pohl et al.~\cite{Pohl2013}, we
  performed experiments to see if the treewidth of graphs of the
  formulas correlates with
  solver running time. We found that for large real-world FMs the
  correlation is weak.

\item We also developed a technique for generating hard artificial FMs
  to better understand the different characteristics between
  easy, large, real-world FMs vs. hard, small, artificial ones.
\end{enumerate}

Prior to our discovery, it was not obvious why SAT-based analysis is so
effective for real-world FMs. Our findings provide strong evidence
supporting the thesis presented in the paper. We believe the
questions we posed are important to better understand the
conditions under which SAT solvers perform efficiently, guiding future
research in technique development. Our findings are especially relevant
for variability aware static analysis, which require making thousands of
SAT queries on the FM. Further, it is possible that SAT solvers do not perform
well on feature models for some specific future real-world problems; the findings
in this paper will be helpful to identify
such challenges and their solutions.
Additionally, we want to emphasize
that all the FMs in our experiments are obtained from
a diverse set of large real-world applications~\cite{BergerVAMOS13,
BergerASE}, including a large FM based off the Linux kernel configuration
model.

\section{Background}
\vspace{0.2cm} This section provides the necessary background on FMs
and the use of SAT solvers in analyzing them.

\subsection{Feature Models (FM)}


Structurally, a FM looks like a tree where each node
represents a distinct feature. The terms node and feature will be
used interchangeably. Child nodes have two flavours: \emph{mandatory}
(the child feature is present if and only if the parent feature is
present) and \emph{optional} (if the parent feature is present then
the child feature is optional, otherwise it is absent). Parent nodes
can also restrict its children with feature groups: \emph{or} (if the
parent feature is present, then at least one of its child features is
present) and \emph{alternative} (if the parent feature is present,
then exactly one of its child features is present). These are the
structural constraints between the child and the
parent.

Structural constraints are often not enough to enforce the integrity
of the model, in which case cross-tree constraints are
necessary. Cross-tree constraints have no restrictions like structural
constraints do, and can apply to any feature regardless of their
position in the tree-part of the model. For this paper, cross-tree
constraints are formulas in propositional logic where features are the
variables in the formulas. Two examples of common cross-tree
constraints are $A \rightarrow B$ (A requires B) and $A \rightarrow \neg B$
(A excludes B).

\begin{table*}
\scriptsize
        \begin{centering}
\begin{tabular}{lrrrrrr}
\hline
Model & Variables & Clauses & Horn (\%) & Anti-Horn (\%) & Binary (\%) & Other (\%) \\\hline
2.6.28.6-icse11 & 6888 & 343944 & 7.54 & 50.80 & 6.19 & 47.29 \\
2.6.32-2var & 60072 & 268223 & 61.81 & 70.66 & 27.02 & 5.01 \\
2.6.33.3-2var & 62482 & 273799 & 63.24 & 70.05 & 27.74 & 5.08 \\
axTLS & 684 & 2155 & 71.04 & 52.99 & 25.85 & 7.19 \\
buildroot & 14910 & 45603 & 76.78 & 61.81 & 40.24 & 2.68 \\
busybox-1.18.0 & 6796 & 17836 & 79.79 & 56.91 & 37.25 & 1.94 \\
coreboot & 12268 & 47091 & 82.67 & 68.53 & 45.73 & 0.59 \\
ecos-icse11 & 1244 & 3146 & 92.59 & 73.36 & 73.27 & 6.64 \\
embtoolkit & 23516 & 180511 & 29.06 & 87.65 & 17.81 & 0.34 \\
fiasco & 1638 & 5228 & 84.87 & 52.85 & 38.49 & 1.42 \\
freebsd-icse11 & 1396 & 62183 & 8.25 & 84.79 & 2.63 & 9.52 \\
freetz & 31012 & 102705 & 76.56 & 52.35 & 34.60 & 3.04 \\
toybox & 544 & 1020 & 90.49 & 67.75 & 46.76 & 0.00 \\
uClinux-config & 11254 & 31637 & 69.23 & 63.42 & 30.73 & 0.96 \\
uClinux & 1850 & 2468 & 100.00 & 75.04 & 50.00 & 0.00 \\
\end{tabular}
                \caption{Clause and variable count of real-world
                  FMs. The last four columns counts the percentage of
                  clauses with the specified property. ``Other" are
                  clauses that are neither Horn, anti-Horn, nor
                  binary. The percentages do not sum to 100\% because
                  Horn, anti-Horn, binary, and other are not
                  disjoint.}
                \label{tab:stats}
        \end{centering}
\end{table*}

\subsection{SAT-based Analysis of Feature Models}
The goal of SAT-based analysis of FMs is to find an assignment to the
features such that the structural and cross-tree constraints are
satisfied. It turns out that there is a natural reduction from feature
models to SAT~\cite{BatorySPLC}. Each feature is mapped to a Boolean
variable, the variable is true/false if the feature is
selected/deselected. The structural and cross-tree constraints are
encoded as propositional logic formulas. The SAT solver can answer
questions like whether the feature model encodes no products.
The solver can also be adapted to handle product
configuration: given a set of features that must be present and
another set of features that must be absent, the solver will find a
product that satsifies the request or answer that no such product
exists. Optimization is also possible such as finding the product with
the highest performance, although for this we need optimization
solvers, multiple calls to a SAT solver, and/or bit-blasting the feature model attributes' integer values
into a Boolean formula. Dead features, features that cannot exist in any valid
products, can also be detected using solvers. More generally, SAT
solvers provide a variety of possibilities for automated analysis of
FMs, where manual analysis may be infeasible. Many
specialized solvers~\cite{FMAlloy,SATConfigurator,FMReasoning} for FM
analysis have been built that use SAT solvers as a backend. It is but
natural to ask why SAT-based analysis tools scale so well and are so
effective in diverse kinds of analysis of large real-world FMs. This
question has been studied with randomly generated FMs based
on realistic parameters~\cite{MendoncaEasySat,BETTY,Pohl2013} where
all the instances were easily solved by a modern SAT solver. In this
paper, we are studying large real-world FMs to explain why they are
easy for SAT solvers.

\section{Experiments and Results}
In this section, we describe the experiments that we conducted to
better understand the effectiveness of SAT-based analysis of FMs. We
assume the reader is familiar with the translation of FMs to Boolean
formulas in conjunctive normal form (CNF), which is explained in many
papers~\cite{BergerVAMOS13, BergerASE}.

\subsection{Experimental Setup and Benchmarks}

All the experiments were performed on 3 different comparable systems
whose specs are as follows: Linux 64 bit machines with 2.8 GHz
processors and 32 GB of RAM.

Table~\ref{tab:stats} lists 15 real-world feature models translated to
CNF from a paper by Berger et al.~\cite{BergerTSE}. The number of
variables in these models range from 544 to 62470, and the number
of clauses range from 1020 to 343944. Three of the models, the
ones named 2.6.*, represent the Linux kernel configuration options for the x86 architecture.
A clause is binary if it contains exactly 2 literals. If
every clause is binary, then the satisfiability problem is called
2-SAT and it is solvable in polynomial time. A clause is
Horn/anti-Horn if it contains at most one positive/negative
literal. If every clause is Horn, then the satisfiability problem is
called Horn-satisfiability and it is also solvable in polynomial time
(likewise for anti-Horn). Binary/Horn/anti-Horn clauses account for
many of the clauses, but not overwhelmingly. Lots of Horn and
anti-Horn clauses does not necessarily imply a problem is easy. For
example, every clause in 3-SAT is either Horn or anti-Horn yet we
currently do not know how to solve 3-SAT efficiently in general.

The real-world models we consider are also very complex at the
syntactic level. For example, 5814 features in the Linux
models declare a feature constraint, which, on average, refers to
three other features. As shown by Berger et al.~\cite{BergerTSE},
these real-world feature models have significantly different
characteristics than those of the randomly generated models used in
previous studies~\cite{MendoncaEasySat, Pohl2013}. In particular,
Mendonca et al.~\cite{MendoncaEasySat} generated models with
cross-tree constraint ratio (CTCR) of maximum 30\%, i.e., the
percentage of features participating in cross-tree constraints. They
also showed that models with higher CTCR tend to be harder. The
real-world feature models we use have much higher CTCR, ranging from
46\% to 96\%~\cite{BergerTSE}.
Hence, given the semantics of these real-world
FMs, not to mention their size, it was not a priori obvious at all
that such models would be easy for modern CDCL solvers to analyze and
solve.

\subsection{Experiment: How easy are real-world FMs}
\begin{figure}
\center
\epsfig{file=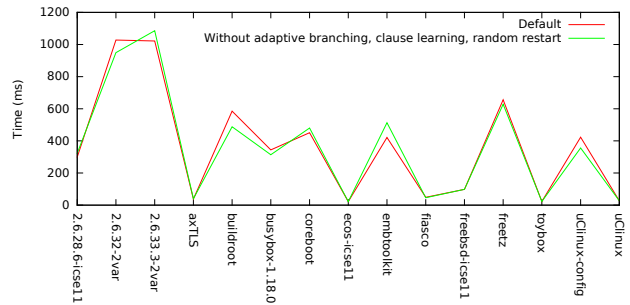,width=\columnwidth}
\caption{Solving times for real-world feature models with Sat4j.}
\label{fig:heurbenchmark}
\end{figure}

Figure~\ref{fig:heurbenchmark} shows the solving times of the
real-world feature models with the Sat4j solver~\cite{sat4j} (version
2.3.4). With the default settings, the hardest feature model took a
little over a second to solve. Clearly, the size of these feature
models is not a problem for Sat4j. Modern SAT solvers have (at least)
4 main features that contribute to their performance: conflict-driven
clause learning with backjumping, random search restarts, Boolean
constraint propagation (BCP) using lazy data structures, and
conflict-based adaptive branching called VSIDS~\cite{Features}. Many
modern solvers also implement pre-processing simplification
techniques. Sat4j implements all these
features except pre-processing simplifications. Figure~\ref{fig:heurbenchmark} shows the running time after
turning off 3 of these features, but the running times did not suffer
significantly. BCP is surprisingly effective for real-world feature
models. The efficiency even without state-of-the-art techniques
suggests that real-world FMs are easy to solve.

\subsection{Experiment: Hard artificial FMs}

The question we posed in this experiment was whether it is possible to
construct hard artificial FMs, and if so what would their structural
features be. Indeed, we were able to construct small feature models
that are very hard for SAT solvers. The procedure we used to generate
such models is as follows:

\begin{enumerate}[leftmargin=*]
\item First, we randomly generate a small and hard CNF formula. One
  such method is to generate random 3-SAT with a clause density of
  4.25. It is well-known that such random instances are hard for a CDCL SAT
  solver to solve~\cite{phasetransition}. We then used such generated problems as the
  cross-tree constraints for our hard FMs.

\item Second, we generate a small tree with only optional
  features. The variables that occur in the cross-tree constraints
  from the first step are the leaves of the tree.
\end{enumerate}

The key idea in generating such FMs is that for any pair of variables
in the cross-tree constraints, the tree does not impose any
constraints between the two. The problem then is as hard as the
cross-tree constraints because a solution for the feature model is a
solution for the cross-tree constraints and vice-versa. Using this
technique, we can create small feature models that are hard to
solve. Unlike these hard artifical FMs, the cross-tree constraints in
large real-world FMs are evidently easy to solve. We also noted that
the proportion of unrestricted variables in hard artificial FMs is
relatively small. 

\subsection{Experiment: Variability in real-world FMs}
Feature models capture variability, and we believe that the search for
valid product configuration is easier as variability increases. We
hypothesized that finding a solution is easy, when the solver has
numerous solutions to choose from. We ran the feature models with
sharpSAT~\cite{SharpSAT}, a tool for counting the exact number of
solutions. The results are in Table~\ref{tab:variability}. We found
that real-world FMs display very high variability, i.e., have lots of
solutions.

\begin{table}
\scriptsize
        \begin{centering}
\begin{tabular}{lrr}
\hline
Model & Variables & Solutions \\\hline
axTLS & 684 & $2^{142}$ \\
busybox-1.18.0 & 6796 & $2^{720}$ \\
coreboot & 12268 & $2^{313}$ \\
ecos-icse11 & 1244 & $2^{418}$ \\
embtoolkit & 23516 & $2^{725}$ \\
fiasco & 1638 & $2^{48}$ \\
freebsd-icse11 & 1396 & $2^{1043}$ \\
toybox & 544 & $2^{57}$ \\
uClinux-config & 11254 & $2^{1388}$ \\
uClinux & 1850 & $2^{303}$ \\ \hline
\end{tabular}
                \caption{The variability in real-world feature models, counted by sharpSAT. The computation finished for 10 of the 15 models.}
                \label{tab:variability}
        \end{centering}
\end{table}

Variability is the reason feature models exist. The feature groups and
optional features increase the variability in the model, and the results
in Table~\ref{tab:variability} suggests the variability grows exponentially with the size of
the model measured in number of variables. High variability in
real-world FMs should have structural properties that make them
easy to solve.

\subsection{Experiment: Why solvers perform very few backtracks for real-world FMs}

The goal of this experiment was to ascertain why solvers make so few
backtracks while solving real-world FMs. First, observe that when a
solver needs to make new decision, it must guess the correct
assignment for the decision variable-in-question under the current
partial assignment in order to avoid backtracking. Also observe that
if a decision variable is unrestricted then either a
{\it true} or a {\it false} assignment to such a variable would lead
to a satisfying assigment. In other words, a preponderance of
unrestricted variable in an input formula to a solver implies that the
solver will likely make few mistakes in assigning values to decision
variables and thus perform very few backtracks during solving.

Given the above line of reasoning, we designed an experiment that
would increase our confidence in our hypothesis that a vast majority
of the variables in real-world FMs are unrestricted, and that the
presence of these large number of unrestricted variables explains why
SAT solvers perform very few backtracks while solving real-world
FMs. The experiment is as follows: whenever the solver branches on a
decision or backtracks (i.e., when the partial assignment changes), we
take a snapshot of the solver's state. We want to examine how many
unassigned variables are unrestricted/restricted. This requires
copying the state of the snapshot into a new solver and checking if
there are indeed solutions in both assignment branches of the variable under the
current partial assignment, in which case the variable-in-question is
unrestricted. If only true branch (resp. false branch) has a solution, then it is a
postively restricted (resp. negatively restricted) variable. If neither branch has a solution, then the current partial
assignment is unsatisfiable.


\begin{figure*}
\centering
\epsfig{file=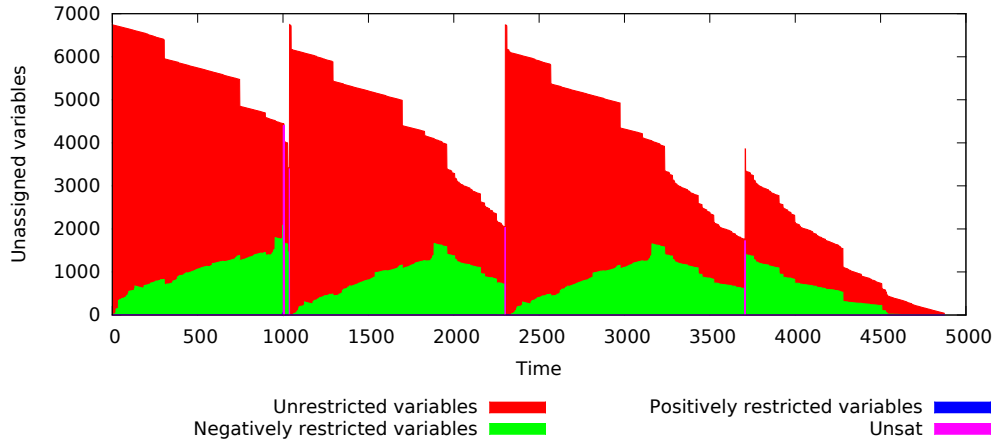}.eps,width=0.75\textwidth}
\caption{Unrestricted and restricted variables for a Linux-based feature model as Sat4j
solves the corresponding Boolean formula. Time is measured by the number of decisions and
conflicts. The y-axis shows the unassigned variables under the current
partial assignment. For example, once the solver reached 500 decisions/conflicts, about
5700 of the variables were unassigned, approximately 1000 of these unassigned variables were negatively restricted variables (green)
and the remainder were unrestricted variables (red). The pink
area means the current partial assignment is unsatisfiable and the
solver will need to backtrack eventually.}
\label{fig:optionalmandatory}
\end{figure*}

Figure~\ref{fig:optionalmandatory} shows how the unrestricted
variables for one Linux-based feature model change over the course of
the search.  The red area denotes unrestricted variables. If
the solver branches on a variable in the red region, the solver will
remain on the right track to finding a solution. The blue/green area
denotes restricted variables. If the solver branches on a variable in
the blue/green region, the solver must assign the variable the correct
value.~\footnote{If the solver picks a random assignment for a
  restricted variable, then the solver will make a correct choice 50\%
  of the time. In practice, SAT solvers bias towards false so
  negatively restricted variables might be preferable to positively
  restricted variables. The bias is often configurable.}  The 15 real-world feature
models have very large red regions.

A large amount of unrestricted variables suggests that the instance is
easy to solve. When the decision heuristic branches on an unrestricted
variable, it does not matter which branch to take, the partial
assignment will remain satisfiable either way. Feature models offer
enough flexibility such that the SAT solvers rarely run into dead
ends.

\subsection{Experiment: Simplifications}
We hypothesized that since the vast majority of the variables are
unrestricted they can be easily simplified away. Furthermore, the
remaining variables/clauses, we call a {\it core}, would be small
enough such that even a brute-force approach could solve it easily. In
point of fact, most, but not all, modern solvers have both pre- and
in-processing simplification techniques already built-in. The term
in-processing refers to simplification techniques that are called in
the inner loop of the SAT solver, whereas pre-processing techniques
are typically called only once at the start of a SAT solving run.

The goal of these experiments was not to suggest a new set of
techniques to analyze FMs, but rather to reinforce our findings for
the effectiveness of SAT-based analysis of real-world FMs.

We implemented a number of standard simplifications that are
particularly suited for eliminating variables. These simplifications
were implemented as a preprocessing step to the Sat4j solver. Note
that the Sat4j solver does not come packaged with variable
elimination techniques, and hence we had to implement them as a
pre-processor. Briefly, a simplification is a transformation on the
Boolean formula where the output formula is smaller and equisatisfiable to the
input formula. We carefully chose simplifications that run in time polynomial
in the size of the input Boolean formula.

What we found was that indeed the simplifications were effective in
eliminating more than 99\% of the variables from real-world FMs. In 11
out of 15 instances from our benchmarks, the simplifications
completely solved the instance without resorting to calling the
backend solver. In the remaining cases, the cores were very small
(at most 53 variables) and were largely Horn clauses that were easily
solved by Sat4j.
The simplifications we implemented are described below:

\vspace{0.5em}
\noindent {\bf Equivalent Variable Substitution:}
If $x \implies y$ and $y \implies x$ where $x$ and $y$ are variables,
then replace $x$ and $y$ with a fresh variable $z$.
The idea behind this technique is to coalesce the variables since
effectively $x=y$. This simplification step is useful for mandatory
features that produce bidirectional implications in the CNF
translation.

\vspace{0.5em}
\noindent {\bf Subsumption:}
If $C_1 \subset C_2$ where $C_1$ and $C_2$ are clauses, then $C_2$ is
subsumed by $C_1$. Remove all subsumed clauses.
The idea here is that subsumed clauses are trivially
redundant. Initially, no clauses are subsumed in the real-world
feature models. After other simplifications, some clauses will become
subsumed.

\vspace{0.5em}
\noindent {\bf Self-Subsuming Resolution:}
If $C \vee x$ and $C \vee \neg x \vee D$
where $C$ and $D$ are clauses and $x$ is a literal, then replace with
$C \vee x$ and $C \vee D$.
The idea here is that if $x=true$ then $C \vee \neg x \vee D$ reduces
to $C \vee D$. If $x=false$ then $C=true$ in which case both $C \vee
\neg x \vee D$ and $C \vee D$ are satisfied. In either case, $C \vee
\neg x \vee D$ is equal to $C \vee D$, hence the clause can be
shortened by removing the variable $x$.

Some features in the 3 Linux models are tristate: include the feature
compiled statically, include the feature as a dynamically loadable
module, or exclude the feature. Tristate features require two Boolean
variables to encode. This simplification step is particularly useful
for tristate feature models. For example, the feature $A$ is encoded
using Boolean variables $a$ and $a'$. Table~\ref{tab:3value} shows how to
interpret the values of this pair of variables.

\begin{table}[htp]
\scriptsize
\center
\begin{tabular}{lll}
\hline
$a$ & $a'$ & Meaning \\\hline
$true$ & $true$ & Include A compiled statically \\
$true$ & $false$ & Include A as a dynamically loadable module \\
$false$ & $false$ & Exclude feature A \\\hline
\end{tabular}
                \caption{Interpretation of tristate features.}
                \label{tab:3value}
\end{table}

To restrict the possibilities to one of these 3 combinations, the
translation adds the clause $a \vee \neg a'$. Since the interpretation
of the feature requires both variables, they often appear together in
clauses. Any other clause that contains $a \vee \neg a'$ will be
removed by subsumption. Any other clause that contains $a \vee a'$ or
$\neg a \vee \neg a'$ will be shortened by self-subsuming resolution.

\vspace{0.5em}
\noindent {\bf Variable Elimination:}
Let $T$ be the set of all clauses that contain a variable and its
negation. Let $x$ be a variable.
$S_x$ is the set of clauses where the variable $x$ appears only
positively. $S_{\overline{x}}$ is the set of clauses where the
variable $x$ appears only negatively. The variable $x$ is eliminated
by replacing the clauses $S_x$ and $S_{\overline{x}}$ with:

$ \{ C_1 \vee C_2\ \mid\ (x \vee C_1) \in S_x,\ (\neg x \vee C_2) \in S_{\overline{x}},\ (C_1 \vee C_2) \notin T \}$

The idea here is to proactively apply an inference rule from propositional logic
called the \emph{resolution} rule. This simplification step is called
variable elimination because the variable $x$ no longer appears in the 
resulting formula. This rule is only applied to variables where the number of output clauses
is less than or equal to the number of input clauses.

This simplification step is very effective for pure literals. A
variable is pure if it appears either only positively or only
negatively in the input formula. If a variable is pure, then variable
elimination will eliminate that variable and every clause containing
that variable. 37.8\% of variables in the 15 real-world feature models
from Table~\ref{tab:stats} are pure. More variables can become pure as
the formula is simplified.

\vspace{0.5em}
\noindent {\bf Asymmetric Branching:}
For a clause $x \vee C$, where $x$ is a literal and $C$ is the
remainder of the clause, temporarily add the constraint $\neg C$. If a
call to BCP returns unsatisfiable, then learn the clause $C$. The new
learnt clause subsumes the original clause $x \vee C$ so remove the
original clause. Otherwise, no changes.

\vspace{0.5em}
\noindent {\bf RCheck:}
For a clause $C$, temporarily replace the clause with the constraint
$\neg C$. If a call to BCP returns unsatisfiable, then the other
clauses imply $C$ so the clause is redundant. Remove $C$ from the
formula. Otherwise, no changes.
The idea is to remove all clauses that are implied modulo BCP. Implied
clauses can be useful for a SAT solver to prune the search space, for
example learnt clauses, but they complicate analysis.

\vspace{0.5em}
\noindent {\bf BCP:}
The last simplification step is to apply BCP: if a clause of length $k$
has $k-1$ of its literals assigned to false, then assign the last literal
to true.

\subsubsection{Fixed Point}
We repeat the simplifications a maximum of 5 times or stop when a fixed
point is reached. The simplifications can only shorten the size of
the input formula with the exception of variable elimination. It is
unclear how many passes of simplification are necessary to reach a fixed
point in the worst-case with variable elimination in the mix. The
upper limit of 5 passes is to guarantee a polynomial (in the size of
the input formula) number of passes. For the models from
Table~\ref{tab:stats}, 2 to 3 passes are enough to reach a fixed
point using the current implementation built on top of MiniSat's
simplification routine. Each additional pass, in practice, experiences
diminishing returns and 5 passes should be sufficient to simplify
real-world feature models. The remaining formula after the 5 passes
of simplifications is the core.

The worst-case execution time of the simplifications is polynomial in
the size of the input formula. We used
a standard encoding of FMs into Boolean formulas that are polynomially
larger than the FMs in the number of features.


\begin{table*}
\scriptsize
        \begin{centering}
\begin{tabular}{lrrrrrr}
\hline
Model & Variables & Clauses & Horn (\%) & Anti-Horn (\%) & Binary (\%) & Other (\%) \\\hline
2.6.28.6-icse11 & 30 & 335 & 93.73 & 6.87 & 93.73 & 0.00 \\
2.6.32-2var & 0 & 0 & NA & NA & NA & NA \\
2.6.33.3-2var & 0 & 0 & NA & NA & NA & NA \\
axTLS & 0 & 0 & NA & NA & NA & NA \\
buildroot & 0 & 0 & NA & NA & NA & NA \\
busybox-1.18.0 & 0 & 0 & NA & NA & NA & NA \\
coreboot & 53 & 153 & 84.31 & 73.86 & 58.17 & 0.00 \\
ecos-icse11 & 0 & 0 & NA & NA & NA & NA \\
embtoolkit & 20 & 67 & 82.09 & 22.39 & 17.91 & 0.00 \\
fiasco & 30 & 384 & 99.74 & 7.03 & 99.74 & 0.00 \\
freebsd-icse11 & 0 & 0 & NA & NA & NA & NA \\
freetz & 0 & 0 & NA & NA & NA & NA \\
toybox & 0 & 0 & NA & NA & NA & NA \\
uClinux-config & 0 & 0 & NA & NA & NA & NA \\
uClinux & 0 & 0 & NA & NA & NA & NA \\\hline
\end{tabular}
                \caption{Clause and variable count of simplified real-world FMs. If the variable count is 0, then the simplification solved the instance because there are no variables remaining.}
                \label{tab:simpstats}
        \end{centering}
\end{table*}

\subsubsection{Simplified Feature Models}
Table~\ref{tab:simpstats} shows the simplified feature models. Simplification alone is
able to solve 11 of the models without resorting to a backend solver. The remaining
4 are shrunk drastically in terms of variables and clauses by simplification.

At least 80\% of the clauses in the simplified models are
Horn. Horn-satisfiability can be solved in polynomial time. The
algorithm works by applying BCP, and the Horn-formula is unsatisfiable
if and only if the empty clause is derived. A similar algorithm for
anti-Horn-satisfiability also exists. BCP is the engine for solving
Horn-satisfiability in polynomial time and modern SAT solvers
implement BCP, hence we hypothesize that Boolean formulas with $\leq
53$ variables and $\geq 80\%$ Horn clauses are easy for SAT solvers to
solve.

\begin{figure}
\center
\epsfig{file=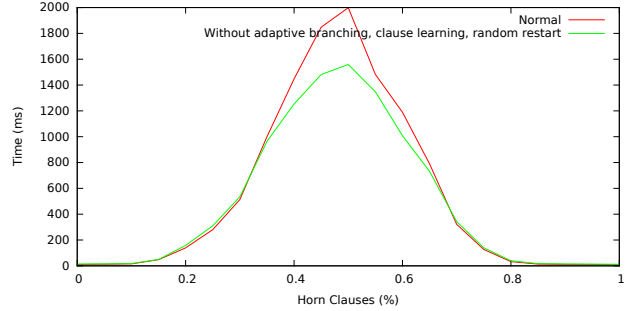,width=\columnwidth}
\caption{Solving Horn clauses for randomized 3-SAT (200 variables and
  850 clauses) with Sat4j. The clause density is set to 4.25, where
  random 3-SAT is hardest to solve.}
\label{fig:horn}
\end{figure}

Figure~\ref{fig:horn} shows how Horn clauses make solving easier. Note
that for 3-SAT, every clause is either Horn xor anti-Horn, hence the
symmetry in the graph. Every point in the graph is the average running
time of 100 randomly generated 3-SAT instances. The instances were
generated with 200 variables, which is larger than the simplified
{\it coreboot} model. The running times where the Horn clauses exceed
80\% is very low. Clause learning, adaptive branching, and random restarts are not necessary
for this case. Although we originally suspected the core might be hard
for its size, the core itself turned out to be easy as well.




\subsection{Experiment: Treewidth of real-world FMs}
\begin{figure}
\begin{centering}
\epsfig{file=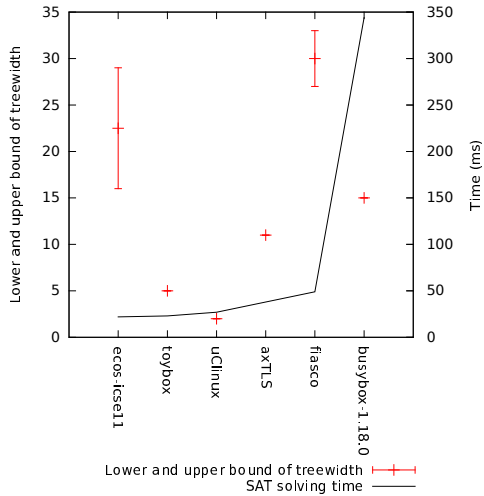, width=0.8\columnwidth}
\caption{Treewidth of real-world FMs plotted against solving time. 4
  models have exact bounds.}
\label{fig:tw}
\end{centering}
\end{figure}

Experiments by Pohl et al.~\cite{Pohl2013} show treewidth of the CNF
representation of randomly-generated FMs to be strongly correlated
with their corresponding solving times. We repeat the experiment on
the real-world FMs to see if the correlation exists. In our
experiments, treewidth is computed by finding a lower and upper bound
because the exact treewidth computation is too expensive to
compute. The longer the calculation runs, the tighter the bounds
are. We used the same algorithm and package used by Pohl et al. We
gave the algorithm a timeout of 3600 seconds and 24 GB of heap memory,
up from 1000 seconds and 12 GB in the original experiment by Pohl et
al. Figure~\ref{fig:tw} shows the results. The computation failed to
place any upper bound on 9 FMs, and we omit these results because we
do not know how close the computed lower bounds are to the exact
answer. For the 6 models in the table, the lower and upper bound are
close, and hence close to the exact answer.

We used Spearman's rank correlation, the same correlation method as in
the original experiment, to correlate the lower bound and time. We
found the correlation between the lower bound and time to be
0.257. We get the same number when correlating between the upper bound and time
(Spearman's rank correlation is based on the rank and the lower and upper
bound rank the models in the same order). Our results show a
significantly poorer correlation than previous work on
randomly-generated FMs. Although our sample is small, research on treewidth
in real-world SAT instances (not FMs) have also found similar
results, i.e., that treewidth of input Boolean formulas is not
strongly correlated with running time of solvers and is not indicative
of an instance's hardness~\cite{MicrosoftTreeWidth}.

\subsection{Interpretation of Results}

It is clear from our experiments that the vast majority of variables
in real-world FMs are unrestricted and can be solved/analyzed by
appropriate simplification and BCP in time polynomial in the number of
variables of the corresponding SAT instance. What remains after BCP and
simplification is a very small set of {\it core} clauses (small relative to the
number of clauses in the input SAT formula) which can be solved by a CDCL solver with very few
backtracks.

\begin{figure}
\center
\epsfig{file=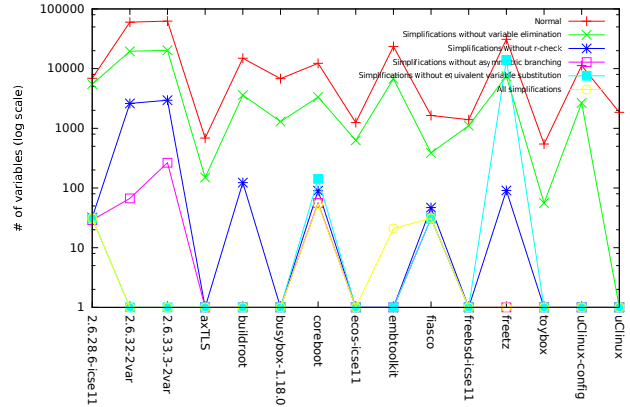, width=\columnwidth}
\caption{The effectiveness of simplifications at reducing the number
   of variables. We try turning off simplifications one at a time and
   measure the result. The line ``normal" is the original input
   Boolean formula with no simplifications.}
\label{fig:simp}
\end{figure}

\section{Threats to Validity of Experimental Methodology and Results}
In this section, we address threats to validity of our experimental
methodology and results.

\vspace{0.5em}
\noindent {\bf Validity of FMs used for the Experiments:} The studied
  collection of FMs are large real-world feature
  models~\cite{BergerTSE}, primarily based on software product lines.
  This collection includes all large real-world models that are
  publicly available. Further, systems software, the domain of these
  models, is known to produce some of the most complex configuration
  constraints~\cite{BergerIST}. After
  taking into account the complexity of these models, we believe that
  these findings will likely hold for many other large real-world
  models.

\vspace{0.5em}
\noindent {\bf Validity of Experimental Methodology:} The unrestricted
  variables experiment takes the partial assignments we encounter
  during the run of the solver with default parameters. The partial
  assignments could vary if the branching heuristic is initiated with a
  random seed. Ideally, we would like to show that unrestricted
  variables are plentiful under any partial assignment but this would
  require an infeasible amount of effort to prove. Having said that,
  there is no reason to believe that the number of unrestricted
  variables would change drastically for a different random seed. In fact, once we
  realized that real-world FMs have large number of unrestricted
  variables, we were able to analytically show that these unrestricted
  variables occur in such formulas due to the high variability of
  FMs. This high variability in FMs is a fundamental property of such
  models, and is independent of translation to SAT or random seed
  chosen.
  

\vspace{0.5em}
\noindent {\bf Correctness of Mapping of FMs into SAT:} The analyses rely on
  the correctness of mapping of the Kconfig and CDL models to
  propositional logic. We rely on existing work that reverse
  engineered the semantics of these languages from documentation and
  configuration tools and specified the semantics
  formally~\cite{BergerASE,BergerTSE}.


\vspace{0.5em}
\noindent {\bf Validity of Choice of Solvers:} All our experiments were
  performed using the Sat4j solver~\cite{sat4j}. An important concern
  is whether these results would apply to other SAT solvers and
  analysis tools built using them. Our choice of Sat4j was driven by
  the fact that it is one of the simplest CDCL SAT solver publicly available,
  and this
  simplicity makes it easier to draw correct conclusions. Note that
  most other SAT solvers used in FM analysis are not only CDCL in the
  same vein as Sat4j, but also implement efficient variants of
  simplification techniques discussed above. Hence, we believe that
  our results would apply equally well to other solvers used in FM
  analysis. Finally, we did not consider CSP or BDD-based solvers since
  CDCL solvers are demonstrably better for FM analysis based on the
  standard encoding used by Benavides et al.~\cite{BenavidesVAMOS}.


\section{Related Work}

Using constraint solvers to analyze feature models has a long
tradition. Benavides et al.~\cite{BenavidesSurvey} give a
comprehensive survey of techniques for analyzing feature models, such
as checking consistency and detecting dead features, and for
supporting configuration, such as propagating feature
selections. These techniques use a range of solvers, including SAT and
CSP solvers. Batory~\cite{BatorySPLC} was first to suggest the use of
SAT solvers to support feature model analyses.

Several authors have investigated the efficiency of different kinds of
solvers for feature model analyses. Benavides et
al.~\cite{BenavidesVAMOS} compare the performance of SAT, CSP, and BDD solvers on
two sample analyses on randomly generate feature models with up to 300
features. They show that BDD solvers are prone to exponential growth
in memory usage on larger models, while SAT solvers achieve good
runtime performance in the experiments.  Pohl et al.~\cite{Pohl2011}
run similar comparison on feature models from the SPLOT collection,
including a wider set of solver implementations. SPLOT models are
relatively small---the largest one has less than 300 features---and
are derived from academic works. Their results show that SAT solvers
work well also for SPLOT models, although they detect some performance
variation for C- vs. Java-based SAT solvers; they also confirm the
tractability challenges for BDD solvers. Mendonca et
al.~\cite{FMCompilationGPCE} achieve scalability of BDD-based analyses
to randomly generated feature models with up to 2000 features by
optimizing the translation from the models to BDDs using variable
ordering heuristics based on the feature hierarchy. Further, Mendonca
et al.~\cite{MendoncaEasySat} show that SAT-based analyses scale to
randomly generated feature models with up to 10000 features. All this
previous work shows that SAT solvers perform well on small realistic
feature models (SPLOT collection) and large randomly generated feature
models. Although SAT solvers have been successfully applied to analyze
the feature model of the Linux kernel~\cite{Undertaker}, we
are unaware of prior work systematically studying the performance of
SAT solvers on large (1000+ features) real-world feature
models, which is what we focus on in this paper.

Previous work has also investigated the hardness of the SAT instances
derived from feature models, aiming at more general insights into the
tractability of SAT-based analyses. Mendonca et
al.~\cite{MendoncaEasySat} have studied the SAT solving behavior of
randomly generated 3-SAT feature models. A 3-SAT feature model is one
whose cross-tree constraints are 3-SAT formulas. While random 3-SAT
formulas become hard when their clause density approaches 4.25
(so-called phase transition), this work shows experimentally that
3-SAT feature models remain easy across all clause densities of their
corresponding 3-SAT formulas. Our work is different since it
investigates the hardness of large real-world feature models rather
than random ones. Even though random feature models are easy on
average, specific instances may still be hard; for example, a tree of
optional features with a hard 3-SAT cross-tree formula over the tree
leaves is, albeit contrived, a hard feature model. Segura et
al.~\cite{BETTY} treat finding hard feature models as an optimization
problem. They apply evolutionary algorithms to generate feature models
of a given size that maximize solving time or memory use. Their
experiments show that relatively small feature models can become
intractable in terms of memory use for BDDs; however, the approach did
not generate feature models that would be hard for SAT solvers. Again,
this work only considers synthetically generated feature models,
rather than real-world ones.

Finally, Pohl et al.~\cite{Pohl2013} propose graph width measures,
such as treewidth, applied to graph representations of the CNF
derived from a feature model as a upper bound of the complexity of
analyses on the model. Their work evaluates this idea on a set of
randomly generated feature models with up to 1000 features using SAT,
CSP, and BDD solvers; they also repeat the experiment on the SPLOT
collection. They find a significant correlation between certain treewidth
measures on the incidence graph and the running time for most of
the solvers. Further, the authors note that it is still unclear why
SPLOT models are easier than generated ones, and whether that
observation would hold for large-scale real-world feature models. Our
work addresses this gap by investigating the hardness of large-scale
real-world feature models and providing an explanation why they are
easy. Moreover, we were not able to identify any correlation between
the treewidth and easiness of the SAT instances derived from large
real-world models, which calls for more work to find effective
hardness measures for feature models.

Large real-world models have become available to researchers only
recently. While some papers hint at the existence of very large models
in industry~\cite{BergerVAMOS13}, these models are typically highly
confidential. Sincero~\cite{Sincero2007} was first to observe that the
definition of the Linux kernel build configuration, expressed in the
Kconfig language, can be viewed as a feature model. Berger et
al.~\cite{BergerASE} identify the Component Definition Language
(CDL) in eCos, an open-source real-time operating system, as
additional feature modeling language and subsequently~\cite{BergerTSE}
create and analyze a collection of large real-world feature models
from twelve open-source projects in the software systems domain. We
use this collection as a basis for our work.



\section{Conclusions}
\vspace{0.2cm} In this paper, we provided an explanation, with strong
experimental support, for the scalability of SAT solvers on large
real-world FMs. The explanation is that the overwhelming majority of
the variables in real-world FMs are {\it unrestricted}, and solvers
tend not to backtrack in the presence of such variables. We argue that
the reason for the presence of large number of unrestricted variables
in real-world FMs has to do with high variability in such models. We
also found that if we switch off all the heuristics in modern SAT
solvers, except Boolean constraint propagation (BCP) and backjumping (no
clause-learning), then the solver does not suffer any deterioration in
performance while solving FMs. Moreover, we ran a set of
simplifications with substantial reduction to the size of the
instances. In fact, a majority of the models were solved
outright with these polynomial-time simplifications. These experiment further bolster our thesis that most
variables in real-world FMs are unrestriced that can be eliminated by
BCP or appropriate simplifications, and do not cause SAT solvers to
perform expensive backtracks. Finally, we note that variables/clauses
that are not eliminated through BCP or simplification, namely the
core, are so few and mostly Horn that backtracking solvers can easily solve them.

Visit the following URL to download further details of the experiments
we
performed and additional associated data: \\
https://github.com/JLiangWaterloo/fmeasy.

\bibliographystyle{abbrv}
\bibliography{sigproc}

\end{document}